\documentstyle[12pt]{article}
\begin{document}
\begin{titlepage}
\pagestyle{empty}
\baselineskip=21pt
\rightline{Alberta Thy-30-92}
\rightline{CERN-TH-6642/92}
\rightline{CfPA-92-27}
\rightline{UMN-TH-1112/92}
\rightline{September 1992}
\vskip .2in
\begin{center}
{\large{\bf ON THE BARYON, LEPTON-FLAVOUR AND  RIGHT-HANDED}}\\
{\large{\bf  ELECTRON ASYMMETRIES OF THE UNIVERSE}}
\end{center}
\vskip .1in
\begin{center}
Bruce A. Campbell,\footnote{ Dept. of Physics, University of Alberta,
Edmonton,
Alberta, Canada T6G 2J1}
Sacha Davidson,\footnote{ Center for Particle
Astrophysics,University of California, Berkeley, California 94720}
John Ellis, \footnote{ CERN, CH-1211 Geneva 23, Switzerland}
and Keith A. Olive \footnote {School of Physics and Astronomy,
University of
Minnesota, Minneapolis, MN 55455}
\end{center}

\vskip .2in

\centerline{ {\bf Abstract} }

\baselineskip=18pt

\noindent
Non-perturbative electroweak effects,
in thermal equilibrium in
the early universe, have the potential
to erase the baryon asymmetry of the
universe, unless it is encoded in a B-L
asymmetry, or in some ``accidentally"
conserved quantity. We first consider the possibility that the BAU
may be regenerated from lepton flavour asymmetries even when
initially $B-L = 0$. We show that provided some, but {\it not} all
the lepton flavours are violated by ${\Delta}L{\neq}0$ interactions
in equilibrium, the BAU may be regenerated without lepton mass
effects.
We next examine the possibility of encoding the baryon asymmetry
in a primordial asymmetry for the right-handed electron,
which due to its weak
Yukawa interaction only comes into chemical equilibrium
as the sphalerons are
falling out of equilibrium.
This would also raise the possibility of preserving an initial baryon
asymmetry when $B-L = 0$.
\vspace*{1cm}

\centerline{ {Published In: {\it Phys.Lett.} {\bf B297} (1992) 118.} }

\vspace*{1cm}

\vfil\eject

\end{titlepage}
    The predominance of matter over antimatter throughout the
observable Universe represents one of the old puzzles of Big Bang
cosmology. It has become more acute in recent years, with the advent
of inflationary models of the development of the early Universe. Any
initial asymmetry would be inflated away, so we would be forced to
rely on the microphysics of elementary particle interactions to
regenerate the Baryon Asymmetry of the Universe (BAU). The basic
requirements for the generation of the BAU were given by Sakharov in
1967 \cite{aaa}.
They are: non-conservation of baryon number, violation of C and
CP symmetry, and out-of-equilibrium dynamics.

    The first realizations of the Sakharov conditions were in the
context of Grand Unified Theories (GUTs). In these theories, quarks
and leptons were unified into multiplets of a larger simple gauge
symmetry, which contained the Standard Model gauge group as a
subgroup. Gauge and Higgs interactions of the larger simple group
then
violated baryon (B) and lepton (L) number at the high energy scales
where grand unification occurred. During the thermal history of the
Universe, heavy gauge and Higgs particles could, via their
out-of-equilibrium decays, generate cosmological baryon and lepton
asymmetries \cite{bb}.
In the simplest GUT models, these asymmetries were
equal, and the net B-L of the Universe was zero. In standard
inflationary cosmology, whilst reheating at the end of the
inflationary epoch seems unlikely to have restored the GUT symmetry,
one can still produce the BAU via the out-of-equilibrium decays of
GUT Higgses, if
they are in turn produced in decays of the quantum of the
inflationary scalar field, the inflaton \cite{cc}.
In supersymmetric models,
the GUT sector can also provide other indirect sources of the BAU, by
inducing low-energy effective operators that introduce
BAU-generating
potential interactions into the coherent oscillations of squark and slepton
v.e.v.'s along flat directions of the scalar potential \cite{dd}.

    The fact that simple GUT models tend to predict that B-L = 0 took
on added importance with the realization that B- and L-violating, but
B-L-conserving, non-perturbative electroweak effects, associated for
example with sphalerons, should be in thermal equilibrium above the
electroweak phase transition temperature $T_C$. These represent a
double-edged source of new physics. First there is the possibility
that
the BAU is actually generated by out-of-equilibrium effects at the
electroweak phase transition, as first analyzed by Kuzmin, Rubakov
and Shaposhnikov \cite{ee}
(KRS1). Whether this is feasible depends sensitively
on the nature of the electroweak phase transition and the sources of
CP violation available in any given model. The possibility is
also opened up of of generating the BAU by electroweak reprocessing
of
an earlier lepton asymmetry, as proposed by Fukugita and Yanagida
\cite{ff} in
the context of the neutrino mass see-saw. On the other hand, should
BAU generation at the electroweak phase transition not occur, the
disquieting prospect arises that any previously-produced BAU would be
erased \cite{arnmac}, if it were not protected by a B-L asymmetry.

    This danger of BAU erasure is pandemic in models with
interactions beyond those of the Standard Model that violate B and/or
L, and do not conserve B-L. This was first noticed by Fukugita and
Yanagida \cite{ff,ff-b}
in the context of the neutrino mass see-saw, where
integrating out the heavy $N_R$ yields an effective $\nu_L \nu_L HH$
dimension-5 interaction. If this interaction were in thermal
equilibrium simultaneously with the sphaleron effects, then B and L
would be relaxed separately to zero. In earlier papers, we and others
\cite{ggg}-\cite{jj}
have extended this argument to constrain many extensions of the
Standard Model, including supersymmetric theories and models yielding
neutron-antineutron oscillations, assuming that the BAU could not be
generated after the electroweak phase transition.

    A possible way that the BAU might be protected, and our limits
avoided, is if there is some other conserved quantum number with a
primordial asymmetry, which cannot be erased by high temperature
equilibrium interactions and can encode the BAU. The most trivial
possibility in the absence of L-violating interactions is B-L, which
is conserved even non-perturbatively in the Standard Model
\cite{ff-a}.
A more
interesting possibility, which was first analyzed in another paper by
Kuzmin, Rubakov and Shaposhnikov \cite{kk}
(KRS2), is that individual lepton
flavour asymmetries in a B-L=0 universe might regenerate a BAU at low
temperatures when
masses in the lepton sector become important.
The criteria for a lepton flavour
asymmetry to yield a BAU via sphaleron effects, set out in KRS2, are:
1) the generation of a lepton flavour asymmetry by the baryogenesis
(or leptogenesis) mechanism
responsible for the generation of the primordial BAU,
2) the absence from equilibrium of any lepton flavour violating
interactions, for at least one of the lepton generations,
3) the presence of mass effects in the lepton sector while the
non-perturbative electroweak processes are in equilibrium. In the
 Standard Model (SM), where
the lepton mass effects are generated after the electroweak phase
transition, this requirement translates into the necessity for the
sphaleron processes to remain in equilibrium below $T_C$, and hence
into a lower bound on the Higgs mass $m_H$.

    If all three KRS2 conditions are satisfied, the equilibrium
condition
is one in which the BAU is (re)generated when the lepton sector
masses switch on. The BAU is then proportional to mass differences in the
lepton
sector. In the case where the lepton
sector masses are those of the leptons alone, KRS2 concluded that
they are probably insufficient to regenerate enough BAU.  The
analysis
of KRS2 has recently been elaborated by Dreiner and Ross \cite{lll},
who reach the
same conclusion and also note that in supersymmetric models slepton
masses might provide mass effects that are more useful for BAU
regeneration.

Following Nelson and Barr \cite{ff-c},
we mentioned the KRS2 loophole in quoting our previous
constraints \cite{ggg},\cite{jj}
on B and L violating interactions, and noted that one
must use lepton flavour asymmetries, since quark-flavour mixing is in
thermal equilibrium. We also quoted two different sets of limits,
considering sphaleron and new interactions either just above $T_C$
or at much higher temperatures. The latter bounds were much stronger
for operator of dimension $d > 4$, but not for $d\leq 4$ operators.
Recently, Ib\'a\~nez and
Quevedo \cite{mm}
have pointed out that chiral gaugino charges inhibit sphaleron
erasure
of baryon and lepton asymmetries in supersymmetric theories
above the effective
supersymmetry-breaking scale.
The effects analyzed by Ib\'a\~nez and Quevedo \cite{mm}
would appear at
high temperature $T \geq {{M_{susy}}^{2/3}}{{M_P}^{1/3}}$, and hence
could only affect the
stronger verions of our bounds for $d > 4$ operators. They do not
affect limits on $d \leq 4$ operators, nor the $T_C$ versions of our
bounds for $d > 4$ operators.

    In this paper we show that while lepton number violating
effective interactions would wipe out a BAU if the interactions
involving all lepton generations were in equilibrium\cite{ggg,hh,jj},
the BAU could actually be (re)generated from lepton flavour
asymmetries
if one or more (but {\it not} all) generations were out of
equilibrium.  We also
examine the possibility that the BAU could be
encoded in a primordial asymmetry for the right-handed electron
field $e_R$,
where the $e_R$ asymmetry arises to balance (part of) the baryon
asymmetry generated by a B-L conserving interaction.

       Let us first consider the (re)generation of a BAU from lepton
flavour asymmetries when initially B-L=0. In particular we wish to
ask if there is a way that this (re)generation can occur, without the
produced BAU depending on the (small) lepton masses. What we will
find is that provided lepton number violation occurs in equilibrium
for some but {\it not} all lepton flavours, then a BAU can be
generated depending on the initial lepton flavour asymmetries (not on
lepton masses). A priori this is unexpected, as equilibrium lepton
number violation for all lepton flavours would inexorably result in
the erasure of all B and all L, while no violation of lepton flavour
would leave us with the situation analyzed in KRS2.

        To illustrate our argument, let us consider the dimension-5
operator  $\nu_L \nu_L HH$, which
violates $L$ (and $B-L$) has been shown \cite{ff-b},
\cite{harv} to lead to the equilibrium condition
$B=L=0$ in the presence of sphaleron interactions
if interactions involving all three generations
are in equilibrium. The requirement that the interactions
be out of equilibrium led to a constraint on the
Majorana mass term which characterizes the strength
of the effective interaction.  Let us write the
additional lagrangian interaction terms as
$h_{ij}HL^i\nu_R^j  +  M_{i}\nu_R^i\nu_R^i$ where we work in a mass
diagonal basis for the $\nu_R^i$, so that the  effective interaction
becomes
\begin{equation}
 \sum\frac{h_{ik} h_{kj}}{M_k} \nu_L^i \nu_L^j H H
\label{xx}
\end{equation}
and the light
neutrino masses given by the see-saw are $m_{\nu_i} \simeq
\sum{h_{ik}h_{ki} v^2 / {M_k}}$ where $v$  is the electroweak
Higgs expectation value.
If the neutrino Yukawa couplings are not
generation-dependent,
one can set a limit on $M' = M/h^2 > 10^{10} - 10^{16}$
GeV
depending on the scale the $B-L$ asymmetry was produced \cite{ff-b},
\cite{harv}.

Consider now however, the possibility that the neutrino
Yukawas are generation-dependent. In this case, it
is quite possible that only one or two generations satisfy the
above bound on $M'$.  In this case a baryon asymmetry could be
(re)generated given an initial
asymmetry in the neutrino generation for which (\ref{xx}) was out of
equilibrium.
To see this, let us assume that for the first two generations the
$h_{1i}$ and $h_{2i}$
are all sufficiently small that
$M'_1$ and $M'_2$ are large enough to
satisfy the
out-of-equilibrium bound. However we assume that some $h_{3i}$ is
large and the rate of the effective interaction
(\ref{xx})  is in equilibrium for the third generation.  Using
\cite{harv}, one can
easily work out the equilibrium conditions on the full set of
chemical
potentials.  We find that the resulting baryon asymmetry can be
expressed as

\begin{equation}
B =  \frac{84}{247} \mu
\label{yy}
\end{equation}
where $\mu$ is the chemical potentials for $(2/3)B - (L_1+L_2)$.
Thus the final asymmetry depends only on an initial
asymmetry in the
first two generations.

       A posteriori this result is less surprising. If initially
B-L=0 and the sphalerons are in equilibrium, then B=0=L, so we can
take this as our initial condition. L=0 with a lepton flavour
asymmetry means that the lepton asymmetry in one generation is
cancelled by that in another lepton generation.If we then introduce a
sufficiently large mass for the $\nu_{\tau}$, we have an L-violating
interaction in equilibrium that modifies the lepton asymmetry stored in
the $\nu_{\tau}$ and $\tau_L$ (which was balancing the lepton number
in the e and $\mu$ generations to give L=0). So in equilibrium we
will no longer have B-L=0 (since now L$\neq$0), and the
nonperturbative electroweak effects will transform some of the lepton
asymmetry into baryons.

       What is surprising about the mechanism is that in the absence
of lepton number violation (for primordial B-L=0) one would have
obtained the lepton mass suppressed BAU calculated in KRS2; with
violation of all lepton flavours the equilibrium condition one would
have obtained  B=0=L. So the efficient generation of a BAU from a
lepton flavour asymmetry requires that some of the lepton flavours
are violated by the lepton number violating operators, and others
are not. Stated differently, for this mechanism of baryogenesis to
occur we get a {\it lower} bound on the strength of the lepton
violation in some generation, as well as an {\it upper} bound on the
strength of the  lepton violation in one of the other generations.

       Clearly we may now extend our analysis to the consideration of
general effective interactions violating lepton number, that we have
analyzed previously for the case when there was no generation
dependence \cite{ggg,jj}. In the new case where one tries to alter
this analysis by baryogenesis from a lepton flavour asymmetry, and
using generation dependence in the effective interaction to erase
some lepton generations but not others, we see that our previous
bounds take on an entirely new aspect. In the new case it is still
true that our bounds on the operators must be satisfied for at least
one of the lepton generations. But to use this baryogenesis mechanism
then implies that the effective interaction must be in equilibrium
for one of the other lepton generations, thus providing a {\it lower}
bound for its strength in this generation. Clearly this new
possibility only applies to operators involving leptons, where there
is no standard model interaction mixing the generations. As such our
previous generation independent analysis is generally valid  for
baryon number violating operators involving only quark fields.

We next turn to the possibilty that an initial baryon asymmetry
is preserved by being stored in $e_R$. This could in principle
occur even with no
$B-L$ nonconservation whatsoever.
Of the 45 chiral states composing the
three fermion generations of the
Standard Model,
the $e_R$ is the most weakly coupled
to the others. It is kept in thermal equilibrium via its coupling
with the hypercharge gauge boson, but this interaction does not
change
its quantum numbers or generation.
The only interaction by which it connects to
the other standard model states is
its Yukawa coupling to the Higgs, which
chirally flips it to $e_L$.
The tiny electron mass tells us that this Yukawa
coupling is particularly weak:
$h_e \simeq {{\sqrt{2}}}m_e/v \simeq 2.1 \times 10^{-6}$.
Comparing the interaction rate (say for H$\leftrightarrow
e_R\bar{e}_L$)
induced by such a coupling at high temperature in the early
universe, with
the expansion rate, we find that such the $e_R$ only comes into
chemical equilibrium at temperatures $\simeq$ 1 TeV. This means that
any
primordially-generated
lepton number that occurred as $e_R$ would be decoupled from
nonperturbative electroweak
effects until this temperature. Since this is close
to the
temperature, at which the sphaleron effects fall
out of equilibrium, it is possible that the $e_R$ may not be
transformed into $e_L$ soon enough for the sphalerons to turn them
into antiquarks, and thereby wipe out the remaining BAU.

We investigate this possibility numerically as a
function of the Standard Model top quark and Higgs masses $m_t, m_H$.
We find that the sphaleron
suppression is
exponentially sensitive to theoretical uncertainties.
With the current best estimates of $T_C$ and the erasure rate
for an $e_R$ asymmetry, it would
wipe out a primordial baryon asymmetry
for reasonable values of $m_t$ and $m_H$.
However,  an increase of $T_C$ by as little as 50\%, or a decrease in
the erasure
rate by
a factor of 3-4, could essentially preserve an
initial $e_R$ asymmetry. This could
occur either as our understanding of the Standard Model
improves, or  in some extension of the Standard
Model. We also update the limits on B- and L-violating interactions
that we derived previously for interactions involving the $e_R$,
assuming the persistence of a primordial BAU.

    We calculate the baryon erasure as the temperature drops towards
the
electroweak phase transition by integrating the rate equations.
It is a two-step process, where first
$e_R$ are chirally flipped to produce an $e_L$ asymmetry,
and then these annihilate against the baryons via sphaleron
interactions.
However, the rate of sphaleron
interaction is so large above threshold, even
with any plausible guess for threshold turn-on,
that effectively it is a one-step process, where the $e_R$ asymmetry
is
flipped and annihilates immediately.
This means that we only need to consider
the rate equation for chirality flip
of the $e_R$ to determine the efficiency of erasure.

The time evolution for the
number density of the $e_R$ asymmetry $\equiv n_R$ is determined by
the difference in the rate equations for $e_R$ and $\bar e_R$. The
rate of change of the $e_R$ number density includes terms due to
Higgs decays and inverse decays, as well as  reactions involving the
hypercharge gauge boson. The gauge interactions always create or destroy
an
$\bar{e_R}$  and an $e_R$ together, so will not
change the $e_R$ asymmetry.
However, this is not the case for the Higgs interactions
which involve an $e_R\bar{e_L}$ or $e_L\bar{e_R}$ pair.
We have checked that scattering processes make insignificant
contributions to the rate equations, and we are justified within the
Standard Model in neglecting CP-violating differences in $H
\leftrightarrow {e_R}\bar{e_L}$ and $\bar{H} \leftrightarrow
{e_L}\bar{e_R}$ rates. Thus the evolution of the
difference $n_R \equiv n_{e_R}-n_{\bar e_R}$ is
\begin{equation}
\frac{dn_R}{dt} = -3Hn_R + 2(n_L - n_R) \Gamma_{ID} + 2n_H \Gamma_D
\label{III}
\end{equation}
where $n_L$ is the asymmetry in the left-handed electrons, $n_H$ is
the asymmetry in the Higgs,and  $\Gamma_D$ and  $\Gamma_{ID} $  are
the thermally averaged  decay and inverse Higgs decay rates.The factor of 2
in equation (\ref{III}) is due to decays (and inverse decays) involving
the charged and neutral Higgs fields.

We assume, for simplicity, that there is no
$H-\bar H$ or $e_L-\bar e_L$ asymmetry. We also neglect final state phase
space factors.
This gives us:
\begin{equation}
\ln \frac{n_R}{n_{R_i}} = - 2\int \Gamma_{ID} dt
\label{IV}
\end{equation}
where we terminate the rate integration when the
temperature falls to $T_o$, where the sphaleron transitions
drop out of equilibrium.
The inverse decay rate $\Gamma_{ID}$ is:
%
\begin{equation}
\Gamma_{ID} = \frac{1}{n_e} \int\frac{d^3{\bf p}}{(2\pi )^32E}
\int\frac{d^3{\bf p'}}{(2\pi )^32E'}
\int\frac{d^3{\bf p}^0}{(2\pi )^32E^0} {(2\pi)^4}
\delta^4 (p+p'-p^0)\vert{\cal M}\vert^2ff'~,
\label{V}
\end{equation}
%
where ${\bf p}$ and ${\bf p'}$ are the $e_R$ and $\bar e_L$ momenta,
$p^0$ is the $H$ momentum, $f$ and $f'$ are the Fermi-Dirac
thermal distributions for the $e_R$ and $\bar e_L$, and the matrix element is
%
\begin{equation}
\vert{\cal M}\vert^2 = 2p.p'~h^2_e
\label{VI}
\end{equation}
%
The inverse decay, $e_R\bar e_L\rightarrow H$ process is related to the
$H\rightarrow \bar ff$ process we considered in \cite{jj},
but has a different numerical coefficient, which we evaluate
to be
%
%
\begin{equation}
\Gamma_{ID} =
\frac{{m_0^3(T)}{h_e^2}I}{12\pi\zeta (3){T^2}}
\simeq \frac{(\ln(2))^2}{24\pi\zeta(3)}\frac{{m_0^2(T)}{h_e^2}}{T}
\simeq 5.3\times 10^{-3}
\frac{m^2_0(T)h^2_e}{T}
\label{VII}
\end{equation}
%
where \cite{fot}:
%
\begin{equation}
I = {\int_1^{\infty}}
{\ln(\frac{\cosh([u+(u^2-1)^{1/2}]{m_o(T)}/4T)}{\cosh([u-(u^2-1)^{1/2
}]{m_o(T)}/4T)})}{\frac{du}{[\exp(\frac{u{m_o(t)}}{T})-1]}}
\label{VIII}
\end{equation}
%
Note that $({m_o(T)}/T)I$ is only weakly dependent on temperature in
the range that we are interested in. Furthermore, the approximation
in equation (\ref{VII}) becomes exact for $m_o(T) \ll T$.
When evaluating $\Gamma_{ID}$, we use the temperature-dependent
effective Higgs mass at zero momentum and zero Higgs vev,
which is known \cite{nn} to be
%
\begin{equation}
m^2_0(T) = 2D(T^2 - {T_o^2})
\label{IX}
\end{equation}
%
where
%
\begin{equation}
D=(2{m_W}^2+{m_Z}^2+2{m_t}^2+{m_H}^2/2)/8v^2~
\label{X}
\end{equation}
and $m_H$ is the zero-temperature Higgs mass.
Calculations indicate that the sphaleron
transitions drop out of equilibrium at:
\begin{equation}
{T_o}^2 = {{T_C}^2}(1 - {E^2}/({{\lambda}_{T_C}}D))
\label{XI}
\end{equation}
where
\begin{equation}
{T_C}^2 =
({m_H}^2-8B{v^2})/4D~,
\label{XII}
\end{equation}
from the standard thermal one-loop calculation \cite{nn},
%
\begin{equation}
B=3(2{m_W}^4+{m_Z}^4-4{m_t}^4)/64{{\pi}^2}{{v}^4}
\label{XIII}
\end{equation}
\begin{equation}
E = \frac{1}{4\pi{v_o^3}}(2{m_W^3} + {m_Z^3})
\label{XIV}
\end{equation}
and
\begin{equation}
{\lambda_{T_{C}}} \simeq \frac{m_H^2}{2v^2}
\label{XV}
\end{equation}
(for the full expression see \cite{nn}); $T_o$ is only very slightly
below $T_C$ for the range of $m_t$ and $m_H$ chosen here.
When combined with the expression for the expansion rate $tT^{2}=C$,
with
$C=3.65\times10^{21}MeV/\sqrt{N}$, and $N=427/4$ for the
Standard Model, the rate (\ref{VII}) gives:
%
\begin{equation}
ln \frac{n_R}{n_{R_i}} = -8.6 \times {{10}^4}GeV \frac{D}{T_o}
\label{XVI}
\end{equation}
%

We wish to emphasize that the erasure suppression factor (\ref{XVI})
is exponentially sensitive to the chirality flip rate, so that
this is one cosmological situation where even a factor of
$\sqrt{2}$ can be important! We also wish to remind the reader that
we have performed our calculation in an approximation where we are
dropping both $n_{\bar{H}}-n_H$ and $n_{\bar e_L }-n_{e_L}$. A full
calculation retaining these terms would require the numerical
integration of Boltzmann equations, and depend on other particle
asymmetries, and will not be attempted here. We expect these
effects to further suppress the final baryon asymmetry. Within the Standard
Model,
estimating the erasure rate $\Gamma$ requires in particular
precise knowledge of $T_C$ and of $m_0(T)$. One cannot yet
be sure that the one-loop estimate (\ref{XII}) of $T_C$ might not be
significantly altered by higher-order effects.
As for $m_0(T)$, we find that the difference in the
suppression factor $n_R/n_{R_i}$ (\ref{XVI}) is typically much
less than a factor of 2
different from what would have been calculated using naively
the zero-temperature value $m_H$. However, in principle the
rate should be calculated using the full dispersion relation
for the Higgs boson as a function of its momentum ${\bf p}^0$.
More uncertainties enter once one goes beyond the Standard
Model. The value of $T_C$ is likely to change
(Giudice \cite{ff-d} has calculated changes in $T_C$ in the
minimal and non-minimal supersymmetric extensions of the
Standard Model), the
Higgs-electron coupling will increase, e.g., in even the
minimal supersymmetric extension of the Standard Model,
the expansion rate of the Universe will change, and there
may be other processes that flip the chirality of the
$e_R$. Therefore, our estimate of the suppression factor
$n/n_i$ should be considered as only qualitative.

In the figure we plot the erasure suppression factor
$n_R/{n_{R_i}}$ (\ref{XVI}) calculated as a
function of $m_t$ for various allowed values of the Higgs mass (as
labelled on the curves).
We can see that the suppression factor
is not small enough to preserve a useful fraction of a
primordial asymmetry stored in the $e_R$. However,
we emphasize again that,
as we see from
equation (\ref{XVI}), the resulting suppression
factor is exponentially sensitive to
both the Higgs mass
and the critical temperature for the electroweak phase transition.

Finally, we consider the implications of
these considerations for cosmological limits on
new interactions beyond the Standard Model.
Previous analyses of the erasure of
a primordial BAU, even with $B-L\neq0$,
depend on  the number densities of all
fields, in which baryon and lepton asymmetries are stored, being
coupled to the thermal soup.
That part of a baryon or lepton asymmetry which is stored
in an uncoupled field will not be
equilibrated, and one can only include the
density it stores once it comes into thermal equilibrium. In the
present
context, the $e_R$ can, in principle,
encode and protect the asymmetry for as
long as it is out of equilibrium, though this requires the mechanism
of
primordial generation to produce at least part of the asymmetry in
the form of the $e_R$.
The non-equilibration limits on operators not
involving the $e_R$ may not be
inferred at temperatures above the $e_R$
equilibration temperature if there is
a primordial $e_R$ asymmetry.
This only affects operators of dimension greater
than five - all others obtain their best limits at
$T_C$ where the $e_R$ is in
equilibrium, and only operators that do not involve the
$e_R$ field - otherwise
their equilibration would bring the $e_R$ into equilibrium.
It also means that for
these operators the best safe limit one can obtain would
be from temperatures of order $T_C$.
These limits are given in \cite{jj}.

In conclusion: we have considered the possibility of (re)generating
the BAU from a lepton flavour asymmetry for primordial B-L=0. We have
found that if in addition to the lepton flavour asymmetry, there is a
lepton number violating interaction that comes into  equilibrium for
at least one lepton generation , but not all lepton generations, then
we may efficiently regenerate the BAU. This yields both upper limits
on the strength of the interaction in at least one lepton generation,
and {\it lower} limits on its strength in some other lepton
generation.
We have also examined the possibility of encoding the baryon
asymmetry of the universe in a ``balancing"
$e_R$ asymmetry, whose incomplete
thermal equilibration would protect the BAU from the depredations of
non-perturbative electroweak baryon number violation.
For the Standard Model,
with the best current estimates of the parameters of
the electroweak phase
transition, the $e_R$ equilibration appears sufficient
to erase a primordial
BAU so encoded, at least for reasonable values of $m_t$ and $m_H$,
but this conclusion is exponentially sensitive to the
parameters of the model, and the dynamics of
the electroweak phase transition.
As such, it is an
issue to be considered in discussions of baryogenesis,
including in
models which have different particle
content near the electroweak
scale.

\vskip .1in

\newpage
\noindent {{\bf Acknowledgements}} \\
\noindent  We wish to thank M. Luty and M. Voloshin for discussions. The work
of
BAC and SD
was supported in part by the Natural Sciences
and Engineering Research council of Canada.
The work of KAO was supported in
part by DOE grant DE-AC02-83ER-40105, and by a
Presidential Young Investigator
Award. BAC and KAO would like to thank the CERN Theory Division for
kind
hospitality during part of this research.

\newpage

\noindent{\bf{Figure Captions}}

\vskip .1in

\noindent Figure 1:  The suppression factor $n_R/{n_{R_i}}$
for the $e_R$
number
density is plotted as a function of $m_t$ for different values of the
Higgs
mass (labelled in GeV).

\end{document}